\documentclass[onecolumn,preprintnumbers,amsmath,amssymb,aps]{revtex4}

\usepackage{graphicx}
\usepackage{amssymb}
\usepackage{epstopdf}
\usepackage{bm}

\begin{document}
    
\title{Nonlocality as Evidence for a Multiverse Cosmology}
 
\author{Frank~J.~Tipler}
\affiliation{Department of Mathematics and Department of Physics, Tulane University, New Orleans, LA 70118}

\date{\today}

\begin{abstract}
I show that observations of quantum nonlocaltiy can be interpreted as purely local phenomena, provided one assumes that the cosmos is a multiverse.  Conversely, the observation of quantum nonlocality can be interpreted as observation evidence for a multiverse cosmology, just as observation of the setting of the Sun can be interpreted as evidence for the Earth's rotation.

\bigskip
\noindent
keyworkds:  multiverse cosmology, nonlocality, entanglement
\end{abstract}

\maketitle

Cosmology is based on general relativity, and relativity is based on the fact that no signal can travel faster than light.  Yet quantum nonlocality appears to transmit information instantaneously.  If one measures one of two electrons in a singlet state to be spin up, one knows with certainty that the other electron will be in a spin down state.  However, had one measured the electron in the horizontal rather than the vertical direction, and measured the spin to be to the left, one would know with certainty the the spin of the other electron would be to the right.  Both would be the case no matter how far separated the electrons were before the decision which way to measure the spin is made.  One electron could be here on Earth, and the other at cosmological distances. it appears that the information of our choice of measurement is being transmitted faster than light.

I shall show that this appearance is due to ignoring the quantum multiverse (\cite{Everett1957}, \cite{DeWittGraham1973}) nature of reality.  Conversely, observation of apparent nonlocality is an experiment to prove the mulitverse nature of the cosmos.  The Aspect experiment is really a cosmological experiment like the Penzias and Wilson experiment.

This fact is not generally appreciated, because the experimenters who conduct nonlocality experiments rarely think that macroscopic objects are subject to quantum mechanics.  Cosmologists know that quantum mechanics applies to everything.  There once was a time when the entire visible universe was the size of an atom, and if quantum mechanics did not apply to the entire universe at that time, what laws of physics did?  But if quantum mechanics applies to the universe at that time, it applies to the universe now,and it applies to smaller regions of the universe now, in particular to human observers and their macroscopic equipment.

Consider an observer about to measure the spin of one electron of a pair with spins in the singlet state.  Let us also suppose that the observer has decided to measure the spin in the vertical direction.  Let us also assume that the experiment is collectively cosmological, by assuming that the experiment has been delayed so as to allow the other electron to move to a cosmological distance: let us suppose that the spin of  the other electron will be measured by an observer in the Andromeda Galaxy, also in the vertical direction. 

At the instant of measurement in the Earth laboratory, the universe splits into two universes, one in which the spin is measured to be spin up, and the other in which the spin is measured to be spin down.  Similarly, in the Andromeda laboratory, the universe splits into two universes, one in which the electron is measured to be be spin down, and the other in which the electron is  measured to be spin down.  The key point is that quantum mechanics forces these measurements to be perfectly correlated: the measured spins of the two electrons will always be in the opposite directions.  This correlation between the laboratories has been transmitted from the Earth laboratory to the Andromeda at a speed slower than light.  The correlation is just what is meant by ``entangled state,'' and the correlation is transmitted by the fact that the interaction of the observer on the electron is also quantum mechanical, and these linear, like all quantum mechanical  interactions.

\begin{figure}
\includegraphics[width=4.0in]{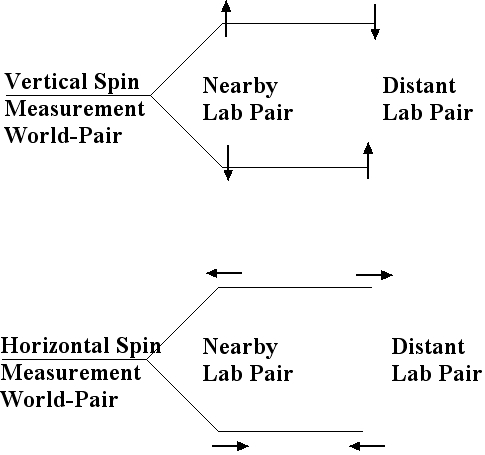}
\caption{\label{fig:fouruniverses}Two universe pairs, one pair in which the spin of the two electrons are to be measured in the vertical direction, and one pair in which the spin of the two electrons are to be measured in the horizontal direction.  In both cases, the spins of the electrons are seen to be perfectly correlated, giving the impression that the first measurement of the spin of one electron has determined the spin of the second  electron. }
\end{figure}

Now suppose the observers had both decided to measure the spins in the horizontal direction.  Once again, the measurement would split the universe into two at each location, but the measurements would be perfectly correlated.  Since the split universes would be perfectly correlated, in reality there would be only four distinct universes, as shown in Figure \ref{fig:fouruniverses}.  	
	
Now let us consider the situation mathematically.  For a pair of electrons, the singlet state can be represented as:

\begin{eqnarray}
|\Psi> = \frac{|\uparrow >_1|\downarrow>_2 - 
 |\downarrow >_1|\uparrow>_2}{\sqrt2} =\nonumber\\
\frac{|\leftarrow >_1|\rightarrow>_2 - |\rightarrow >_1|\leftarrow>_2}{ \sqrt2}\nonumber\\
\label{eq:electronsinglet}
\end{eqnarray}
\noindent
where the former expression is the appropriate basis if the observers wish to measure spin up or down, and the latter basis if measuring the spin horizontally.  $|\uparrow >_i$ or  $|\downarrow >_i$  means the ith electron is spin up and down respectively.  

As described above, the impression of nonlocality arises because we know that if we measure the spin of electron 1, and find it to be spin up, we know for certain that the spin of electron will be spin down, without measuring it, and no matter how far electron 2 is from electron 1  when we measure 1.  Let us now show that nonlocality disappears when we apply quantum mechanics to the observers as well.  I shall use the standard Many-Worlds notation for the observer as presented in \cite{DeWittGraham1973}.

Let $|\ldots>_i|M_i(\ldots)>$ be the composite state of the ith electron and the apparatus that will measure it bebore the measurement, and $U_i$ the operator that represents the measurement interaction, resulting in

\begin{equation}
U_i|\uparrow>_i|M(\ldots)_i> = |\uparrow>_i|M_i(\uparrow)> 
\label{eq:measureelectroni}
\end{equation}

\noindent
for either electron i, and for any $|\uparrow>$, $|\downarrow>$, $|\leftarrow>$, $|\rightarrow>$.

The crucial point is that for the entangled state( \ref{eq:electronsinglet}), the interaction $U_i$ does not result in a single electron state for either electron, but rather it results in a linear superposition, with the measuring apparatus now entangled with the two electrons:

\begin{eqnarray}
U_1\left(\frac{|\uparrow >_1|\downarrow>_2 - 
 |\downarrow >_1|\uparrow>_2}{\sqrt2)}\right)|M_1(\ldots)> =\nonumber\\
 \frac{1}{\sqrt2}\left(|\uparrow >_1|\downarrow>_2M_1(\uparrow)\right) -
\frac{1}{\sqrt2}\left|\downarrow>_1|\uparrow>_2M_1(\downarrow)\right)
\label{eq:splitelectron1}
\end{eqnarray}

In other words, the universe has now been split into two worlds, one in which electron 1 has spin up, and is measured to have spin up, and another world in which electron one has spin down, and has been measured to have spin down.  Electron 2, having not yet been measured, is neither.  

However, notice in (\ref{eq:splitelectron1}) that electron 2 has its spin down state correlated uniquely with the world in which electron 1 has spin up, and its spin down state correlated uniquely with the world in which electron 1 has spin down.  This means that should we measure the spin of electron 2, the result of the measurement will match the result in the appropriate world of electron 1.  That is, a measurement of electron 2 will split the universe into two worlds also, but the previous entanglement will force the two separate splits to be perfectly correlated.  If electrons 1 and 2 are far separated before the measurements on each, the correlations are still carried --- at a speed slower than light --- and it is this correlation, NOT nonlocality, that forces us to observe the two electron spins to be correlated.   If we had instead made the choice to measure the spins in the left-right  rather than up-down, the multiverse would have split into two worlds at each electron measurement location, and as before, the entanglement forces the splits to be correlated.  So as before, the world in which electron 1 is seen to have spin left is necessarily the world in which electron 2 is seen to have spin right and the world in which  electron 1 is seen to have spin right is necessarily the world in which electron 2 is seen to have spin left.

Like the electrons, and like the measuring apparatus, we are also split when we read the result of the measurement, and once again our own split follows the initial electron entanglement.  Thus quantum nonlocality does not exist.  It is only an illusion caused by a refusal to apply quantum mechanics to the macroworld, in particular to ourselves.

Many-Worlds quantum mechanics, like classical mechanics is completely deterministic.  So the observers have only the illusion of being free to chose the direction of spin measurement.  However, we know my experience that there are universes of the mutilverse in which the spins are measured in the orthogonal directions, and indeed universes in which the pair of directions are at angles $\theta$ at many values between $0$ and $\pi/2$ radians.  To obtain the Bell Theorem quantum prediction in this more general case, where there will be a certain fraction with spin in one direction, and the remaining fraction in the other, requires using Everett's assumption that the square of the modulus of the wave function measures the density of universes in the multiverse.  The calculation is more complicated, and will be published elsewhere.  In the general case, as in the horizontal-vertical special case, all measurements are entirely local.  The distant system does not know what has happened far away.  It is only that ignoring the existence of the multiverse structure of the cosmos has given us the illusion of nonlocal effects.  The multiverse is the true hidden variable, and this hidden variable is entirely local.

Have you ever seen the Earth rotate on its axis.  I have.  I have seen the Sun set.  We know today that in spite of our language, it is the Earth that is moving, and not the Sun.  Copernicus, in his {\it On the Revolutions} argued that it was implausible to think the Sun, known at the time to be much larger than the Earth, was the moving body.  Today, it is implausible that there can be nonlocal effects.  Instead, apparently nonlocal effects are really local effects in which we do not see the locality in action.  This is how inflation theory explains the observed fact that the CMBR temperature is observed to be the same on opposite sides of the sky: the two sides were once in causal contact.  

The multiverse conceptual revolution, initiated by Everett, is as fundamentally revolutionary as the Copernican Revolution.  It is essential to understanding quantum reality, which is true reality.  I have shown in this paper that we have actually seen the mulitverse, though we have not realized it: when ever we see quantum nonlocality, we are actually seein a manifestation of the multiverse.

\end{document}